\documentclass[10pt,aps,prd,preprintnumbers,groupedaddress,showpacs,showkeys]{revtex4-2}
\usepackage{amsmath,amssymb,graphicx,epsfig,color,xcolor,ulem,cancel,datetime,dcolumn,bm,mathrsfs,lineno}

\newcommand{\nn}{\nonumber}
\newcommand{\dsf}{\mathrm{dS}_{4}}
\newcommand{\eff}{\mathrm{eff}}
\newcommand{\x}{\mathbf{x}}
\newcommand{\Leff}{\mathcal{L}_{\eff}}
\newcommand{\Lren}{\mathcal{L}_{\mathrm{ren}}}
\newcommand{\Lct}{\mathcal{L}_{\mathrm{c.t.}}}
\newcommand{\Lone}{\mathcal{L}_{1}}
\newcommand{\Ltwo}{\mathcal{L}_{2}}
\newcommand{\csch}{\mathrm{csch}}
\newcommand{\divp}{\mathbf{divp}}
\newcommand{\linp}{\mathbf{linp}}
\newcommand{\Lsqed}{\mathcal{L}_{\mathrm{sQED}}}
\newcommand{\Lmax}{\mathcal{L}_{\mathrm{Maxwell}}}
\newcommand{\LEH}{\mathcal{L}_{\mathrm{EH}}}
\newcommand{\Ltot}{\mathcal{L}_{\mathrm{tot}}}
\newcommand{\B}{\bar{B}}
\newcommand{\V}{\mathcal{V}}
\newcommand{\GeV}{\mathrm{GeV}}
\newcommand{\reh}{\mathrm{reh}}
\newcommand{\G}{\mathrm{Gauss}}
\begin{document}
\title{Scalar QED effective action in dS: implications for primordial magnetogenesis and the running of coupling constants}
\author{Ehsan Bavarsad}
\email{bavarsad@kashanu.ac.ir}
\affiliation{Department of Physics, University of Kashan, 8731753153, Kashan, Iran}
\author{Sang Pyo Kim}
\affiliation{Center for Relativistic Laser Science, Institute for Basic Science, Gwangju 61005, Korea}
\affiliation{Institute of Fundamental Physics and Quantum Technology and School of Physical Science and Technology, Ningbo University, Ningbo, Zhejiang 315211, China}
\affiliation{ICRANet, Piazzale della Repubblica 10, 65122 Pescara, Italy}
\affiliation{School of Advanced Science and Technology, Kunsan National University, Kunsan 54150, Korea}
\author{Cl\'{e}ment Stahl}
\affiliation{Universit\'{e} de Strasbourg, CNRS, Observatoire astronomique de Strasbourg, UMR 7550, 67000 Strasbourg, France}
\author{She-Sheng Xue}
\affiliation{ICRANet Piazzale della Repubblica, 10-65122, Pescara, Italy}
\affiliation{Physics Department, Sapienza University of Rome, Rome, Italy}
\affiliation{INFN, Sezione di Perugia, Perugia, Italy}
\affiliation{ICTP-AP, University of Chinese Academy of Sciences, Beijing, China}
\begin{abstract}
Particles creation and vacuum polarization under the influence of both an electromagnetic field and a de~Sitter (dS) spacetime nonperturbatively probes quantum electrodynamics (QED) and quantum gravitational effects. By applying the gamma-function regularization to the in-out formulation, we find the exact one-loop effective action in the proper-time integral representation for a charged scalar field in a uniform electric field and a parallel magnetic field in a dS spacetime. It reduces to the one-loop scalar QED Weisskopf-Schwinger effective action in the limit of Minkowski spacetime and the one-loop effective action in the pure dS spacetime. After carefully considering the different limiting cases, a striking result concerns the effective strengths of the coupling constants in the pure strong electric field limit. Our study provides further evidence for evolution of Newton's gravitational constant under the influence of the external fields. Remarkably, the analysis of the effective potential shows that the leading order contributions to the vacuum polarization amplitude, depending on the nonminimal coupling threshold, give a nonzero vacuum expectation value of the magnetic field. This mechanism generates a large magnetic field in the early dS stage of the expansion that may imply primordial magnetogenesis.
\end{abstract}
\maketitle
\section{\label{sec:intro}Introduction}
Both a background gauge field or curved spacetime polarizes the vacuum through quantum fluctuations and gives an effective action to the underlying classical theory. The quantum electrodynamics action in a constant electromagnetic field, found by Heisenberg-Euler, Weisskopf, and Schwinger, is the most prominent one-loop action, which is equivalent to summing all one-loop Feynman diagrams with external photons interacting with virtual particles from the Dirac sea \cite{Heisenberg:1936nmg,Weisskopf:1936hya,Schwinger:1951nm}. The real part of the one-loop effective action makes the Maxwell theory nonlinear and results in the vacuum polarization. The imaginary part leads to the vacuum decay due to the so-called (Sauter-) Schwinger pair production in a strong electric field \cite{Schwinger:1951nm,Sauter:1931zz}. Thus, the one-loop QED action exhibits many interesting phenomena, such as the vacuum birefringence and photon splitting \cite{Adler:1971wn}, needless to say the Schwinger pair production in strong electric fields \cite{Dunne:2008kc,Gelis:2015kya,Fedotov:2022ely}. The physics of QED in strong electromagnetic fields has been extensively studied (see, e.g.,~Ref.~\cite{Kranas:2025dwz} for a recent example on its relation to entanglement), it is expected to be detected using ultra-intense lasers \cite{DiPiazza:2011tq}, and is also considered in astrophysics \cite{Harding:2006qn,Ruffini:2009hg}. In the context of condensed matter physics, the Schwinger-like effect has been observed in graphene \cite{Berdyugin:2021njg,Schmitt:2022pkd}.
\par
A dS spacetime emits Gibbons-Hawking radiation of all species of particles from the cosmological horizon \cite{Gibbons:1977mu}. The interplay of gravitational and electromagnetic effects is an attractive arena to study various limiting cases, strong coupling between two effects, and possible physical setups for applications \cite{Chernodub:2016lbo,Jones:2018uad}. An electric field produces pairs of charged particles and antiparticles and a dS spacetime also creates charged pairs, so a uniform electric field in the dS spacetime creates charged pairs whose distribution exhibits the intertwinement of both the Schwinger effect and Hawking radiation \cite{Garriga:1994bm,Kim:2008xv,Garriga:2012qp}. The one-loop scalar QED effective action in the two-dimensional dS and anti-de~Sitter (AdS) spacetimes has been obtained in Ref.~\cite{Cai:2014qba}. Following Ref.~\cite{Bavarsad:2017oyv}, we shall consider a charged scalar field in a uniform electromagnetic field background in the four-dimensional de~Sitter ($\dsf$) spacetime with the metric
\begin{equation}\label{metric}
ds^{2}=\Omega^{2}(\tau)\Big(d\tau^{2}-dx^{2}-dy^{2}-dz^{2}\Big),
\end{equation}
where $\tau\in(-\infty,0)$ is the conformal time, $\x=(x,y,z)$ are coordinates on the flat three-dimensional Euclidean space $\mathbb{R}^{3}$, and the conformal scale factor with the Hubble constant $H$ is expressed as
\begin{equation}\label{scale}
\Omega(\tau)=-\frac{1}{H\tau}.
\end{equation}
We assume the uniform electric and magnetic field backgrounds are parallel to each other, which can be given by the electromagnetic vector potential
\begin{equation}\label{vector}
A_{\mu}(\tau,\x)=By\delta_{\mu}^{1}-\frac{E}{H^2\tau}\delta^{3}_{\mu},
\end{equation}
where $E$ and $B$ are constants denoting the field strength. A comoving observer with four-velocity $u_{\mu}=(\Omega,0,0,0)$ will find the electric and magnetic four-vector fields as
\begin{align}\label{fourv}
E_{\mu}&=u^{\nu}F_{\mu\nu}, & B_{\mu}&=\frac{1}{2}\epsilon_{\mu\rho\nu\sigma}u^{\rho}F^{\nu\sigma},
\end{align}
where $F_{\mu\nu}$ is the electromagnetic field tensor, and the Levi-Civita tensor is defined as $\epsilon_{0123}=-\sqrt{-g}$ in which $g$ is the determinant of the metric. Then, such an observer has
\begin{align}\label{EBvector}
E_{\mu}&=-E\Omega(\tau)\delta_{\mu}^{3}, & B_{\mu}&=-B\Omega^{-1}(\tau)\delta_{\mu}^{3},
\end{align}
and measures the electromagnetic energy density
\begin{equation}\label{emenergy}
-\frac{1}{2}\Big(E_{\mu}E^{\mu}+B_{\mu}B^{\mu}\Big)=\frac{1}{2}\Big(E^{2}+B^{2}\Omega^{-4}(\tau)\Big).
\end{equation}
Hence in the comoving frame the electric field contribution to the electromagnetic energy density is constant through the spacetime, whereas the contribution of the magnetic field decreases with the expansion of $\dsf$ spacetime as $\Omega^{-4}(\tau)$. Notice that the amplitude of the magnetic field, i.e., $\sqrt{|B_{\mu}B^{\mu}|}$ falls off as $\Omega^{-2}(\tau)$, whose behaviour is required by the flux conservation \cite{Durrer:2013pga,Subramanian:2015lua}. We denote the redshifted magnetic field amplitude $B\Omega^{-2}(\tau)$ by the symbol $\B$. The interaction of the complex scalar field $\varphi(x)$ of mass $m$ and the gauge coupling constant $e$ with the electromagnetic field (\ref{vector}) is described by the action
\begin{equation}\label{action:sc}
S=\int d^{4}x\sqrt{-g}\Big\{g^{\mu\nu}\big(\partial_{\mu}+ieA_{\mu}\big)\varphi\big(\partial_{\nu}-ieA_{\nu}\big)\varphi^{\ast}
-\big(m^{2}+\xi R\big)\varphi\varphi^{\ast}\Big\},
\end{equation}
where $\xi$ is a dimensionless nonminimal coupling to the scalar curvature $R=12H^{2}$ of the $\dsf$ spacetime. The Schwinger effect due to the electromagnetic background (\ref{vector}) is given by the production rate of the charged spinless pairs as \cite{Bavarsad:2017oyv,Bavarsad:2018lvn}
\begin{equation}\label{rate}
\Gamma=\Big(\frac{e\B}{2\pi}\Big)\Big(\frac{H^{2}\gamma}{2\pi}\Big)\Big(\frac{1}{e^{4\pi\gamma}-1}\Big)
\bigg[\frac{1}{2}+\sum_{n=0}^{\infty}e^{2\pi(\gamma +\bar{\kappa}_{n})}\bigg],
\end{equation}
written in terms of the convenient dimensionless parameters
\begin{align}\label{lambda}
\mu&=\frac{m}{H}, & \lambda&=\frac{eE}{H^{2}},
\end{align}
the quantities $\gamma$ and $\bar{\kappa}_{n}$ are expressed as
\begin{align}\label{gamma}
\gamma&=\sqrt{\lambda^{2}+\mu^{2}+12\bar{\xi}-\frac{1}{4}}, & \bar{\kappa}_{n}&=\frac{\lambda\gamma}{\sqrt{\gamma^{2}+(2n+1)e\B/H^{2}}},
\end{align}
where $\bar{\xi}=\xi-1/6$ is null in the conformally coupled case. The sum in Eq.~(\ref{rate}) is taken over the quantum number $n$ numbering the Landau levels of the charged scalar field in the magnetic field background. The Schwinger formula reduces to scalar QED effective action in the Minkowski spacetime in the limit $H =0$, and Hawking radiation in the limit of zero electromagnetic field \cite{Bavarsad:2017oyv}. Quantum electrodynamics in curved spacetimes unveils interesting physics and intriguing features of quantum field theory. For instance, the Schwinger effect in a uniform electric field in two-dimensional dS spacetime has a thermal interpretation for the leading Boltzmann factor by an effective temperature of the generalized mean of exponent two between the Hawking temperature and the Unruh temperature of the accelerating charge due to the electric field \cite{Cai:2014qba}. The Schwinger effect \cite{Garriga:1994bm,Kim:2008xv,Kim:2014iba,Sharma:2017ivh,Geng:2017zad,Rajeev:2019okd,Baloi:2019wgd,Chen:2025xrv}, the induced current \cite{Frob:2014zka,
Kobayashi:2014zza,Bavarsad:2016cxh,Hayashinaka:2016dnt,Stahl:2015gaa,Stahl:2016geq,Hayashinaka:2016qqn,Hayashinaka:2018amz,Bastero-Gil:2025jio,
Bastero-Gil:2025nfi,Botshekananfarda:2026lpi}, and the induced energy-momentum tensor \cite{Meimanat:2023hjq,AkbariAhmadmahmoudi:2021tpj,
Botshekananfard:2019zak,Bavarsad:2019jlg,Bavarsad:2018pvc,Bavarsad:2017wbe} are manifestations of pairs production due to background uniform electric field in the dS spacetime and have been intensively investigated. The induced current through the Schwinger effect, which reflects the backreaction of a charged quantum field to both the background electric field and curved spacetime, crucially depends on renormalization schemes \cite{Ferreiro:2018qzr}. An intriguing feature of the Schwinger effect is the infrared (IR-)hyperconductivity, in which the induced current increases though the electric field decreases in a certain range of the field strength \cite{Frob:2014zka,Kobayashi:2014zza,Bavarsad:2016cxh,Hayashinaka:2016dnt}. The IR-hyperconductivity is removed in a renormalization scheme of maximal subtraction, and the negative induced current is explained by analogy with the Hawking radiation \cite{Hayashinaka:2018amz}; see also \cite{Bastero-Gil:2025nfi}. It is argued in Ref.~\cite{Banyeres:2018aax} that the IR-hyperconductivity is due to the nonlinear terms of the Maxwell theory, motivating a more detailed study of the one-loop effective action. The Schwinger effect may lead to some physical applications; for instance, it has been used to model bubble nucleation, \cite{Frob:2014zka,Garriga:1993fh}, to better model primordial magnetogenesis \cite{Kobayashi:2014zza,Banyeres:2018aax,Stahl:2018idd,Sobol:2018djj,Sobol:2019xls,Kobayashi:2019uqs}, to generate primordial nongaussianities \cite{Chua:2018dqh}, and to produce dark matter \cite{Arvanitaki:2021qlj,Bastero-Gil:2023mxm}. Whether the assumption of a constant electromagnetic
field in dS spacetime is a relevant setup is still being debated \cite{Giovannini:2018qbq,Kitamoto:2018htg,Shakeri:2019mnt}. The Schwinger mechanism has also been considered in axion inflation \cite{Domcke:2022vhk,vonEckardstein:2024tix,Iarygina:2025ncl}, for holographic membrane nucleation \cite{Arcos:2022icf}, and with a stochastic \cite{VicenteGarcia-Consuegra:2025lkh,Garcia-Consuegra:2026snp} or a quantum kinetic approach \cite{Lysenko:2023wrs}. All these works motivate the pursuit of the investigation of the Schwinger effect, in particular, the one-loop scalar QED effective action in the $\dsf$ spacetime.
\par
The main purpose of this paper is to find the one-loop scalar QED effective action in the proper-time representation in the $\dsf$ spacetime (\ref{metric}) in the presence of the electromagnetic field background (\ref{vector}). Following up on the results of \cite{Banyeres:2018aax}, we employ another method, the in-out formalism based on the Schwinger action principle \cite{DeWitt:1975ys,Gabriel:2000mg,DeWitt:2003pm}, which yields the one-loop effective action as
\begin{equation}\label{principle}
S_{\eff}=\pm i\sum_{\textbf{K}}\log\big(\alpha^*_{\textbf{K}}\big),
\end{equation}
where $\alpha_{\textbf{K}}$ is the Bogoliubov coefficient with the quantum numbers $\textbf{K}$ and the plus (minus) sign is for bosons (fermions), and then apply the gamma-function regularization to express the one-loop effective action in the proper-time integral representation \cite{Kim:2009pg,Kim:2008yt}. The gamma-function regularization leads to a complex effective action when particles or pairs are produced due to background fields, whose real part is the vacuum polarization and twice of whose imaginary part is the vacuum persistence amplitude \cite{DeWitt:1975ys}. In fact, the vacuum persistence amplitude is a consequence of spontaneous pair production and in the in-out formalism is the sum of all residues at simple poles of the complex effective action according to the Cauchy residue theorem and is determined by the mean number of created pairs
\begin{equation}\label{Im:Seff}
2\Im S_{\eff}=\pm\sum_{\mathbf{K}}\log\big(1\pm\mathcal{N}_{\mathbf{K}}\big),
\end{equation}
the upper (lower) sign is for boson (fermion) \cite{Kim:2008yt,PauchyHwang:2009rz}. Thus the gamma-function regularization will give the one-loop scalar QED effective action consistent with the Schwinger effect formula (\ref{rate}) derived in Refs.~\cite{Bavarsad:2017oyv,Bavarsad:2018lvn}. Our one-loop effective action in the proper-time integral representation is different from the perturbative action in a curved spacetime from the worldline formalism \cite{Dalvit:2000ay,Bastianelli:2008cu,Davila:2009vt}. The one-loop effective action in a pure dS spacetime was found in
Refs.~\cite{Candelas:1975du,Das:2006wg,Kim:2010cb,Akhmedov:2019esv,Akhmedov:2024axn,Jiang:2020evx,Zhou:2025jwm,Chen:2026trv}.
\par
The organization of this paper is as follows.
In Sec.~\ref{sec:persi}, we find the Bogoliubov coefficients between the in- and out-vacuum states for a charged scalar field in the uniform electromagnetic field background (\ref{vector}) in the $\dsf$ spacetime (\ref{metric}), and then employ the zeta-function regulation method to find the exact one-loop effective action in the proper-time integral representation from the scattering amplitude between the in- and out-vacuum states. Furthermore, we check the consistency of the vacuum persistence and the mean number of produced pairs. In Sec.~\ref{sec:polar}, we discuss the renormalization of the real part of the effective Lagrangian in both electric and magnetic field backgrounds, and compute it for subcritical electric and magnetic fields in the weak curvature condition. In Sec.~\ref{sec:strong}, the effective Lagrangian in the pure strong magnetic and electric backgrounds is separately investigated, and its physical implications for magnetogenesis and the running of the coupling constants are explored. Finally, the main conclusions of this study are presented in Sec.~\ref{sec:concl}. The Appendix consists of some of the mathematical formulas that may be useful to make the technical details more transparent.
\section{\label{sec:persi}vacuum persistence in parallel electric and magnetic fields}
The Euler-Lagrange equation of the action (\ref{action:sc}) for the complex scalar field $\varphi(x)$ interacting with the electromagnetic field background (\ref{vector}) in the $\dsf$ spacetime (\ref{metric}) yields the Klein-Gordon equation
\begin{equation}\label{eq:KG}
\bigg[\frac{\partial^{2}}{\partial\tau^{2}}+2H\Omega(\tau)\frac{\partial}{\partial\tau}+H_{\perp}({\x}_{\perp})
-\Big(\frac{\partial}{\partial z}+ieA_{3}(\tau)\Big)^{2}+\Big(m^{2}+12\xi H^{2}\Big)\Omega^{2}(\tau)\bigg]\varphi\big(\tau,\x_{\perp},z\big)=0,
\end{equation}
where the interaction Hamiltonian operator on the transverse plane $\x_{\perp}=(x,y)$ which is perpendicular to the magnetic field background reads
\begin{equation}\label{Hamilt:B}
H_{\perp}({\x}_{\perp})=-\frac{\partial^{2}}{\partial y^{2}}-\Big(\frac{\partial}{\partial x}+ieBy\Big)^{2}.
\end{equation}
The Hamiltonian (\ref{Hamilt:B}) has the normalized eigenfunctions
\begin{align}\label{hermit}
H_{n,k_{x}}^{\pm}({\x}_{\perp})&=\bigg(\sqrt{\frac{eB}{\pi}}\frac{1}{2^{n}n!}\bigg)^{\frac{1}{2}}
\exp\Big(\pm ixk_{x}-\frac{\bar{y}_{\pm}^{2}}{2}\Big)H_{n}(\bar{y}_{\pm}), & \bar{y}_{\pm}&=\sqrt{eB}y\pm\frac{k_{x}}{\sqrt{eB}},
\end{align}
where the upper (lower) sign is for the positive (negative) frequency mode, with $H_{n}$ being the Hermite polynomial, which has the Landau levels $\varepsilon_{n}=(2n+1)eB$ for $n=0,1,2,\ldots$. The separation of variables is accomplished by the substitution
\begin{equation}\label{substit}
\varphi^{\pm}\big(\tau,\x_{\perp},z\big)=\Omega^{-1}(\tau)e^{\pm izk_{z}}f^{\pm}(\tau)H^{\pm}_{n,k_{x}}({\x}_{\perp}),
\end{equation}
into Eq.~(\ref{eq:KG}). Using,
\begin{equation}\label{kn}
k_{n}=\sqrt{k_{z}^{2}+(2n+1)eB},
\end{equation}
the differential equation for $f^{\pm}(\tau)$ is customarily expressed in terms of the dimensionless variable $\bar{\tau}_{\pm}=\pm2ik_{n}\tau$ as
\begin{equation}\label{Whittaker}
\frac{d^{2}f^{\pm}}{d\bar{\tau}_{\pm}^{2}}+\bigg(\frac{-1}{4}+\frac{i\kappa_{n}}{\bar{\tau}_{\pm}}
+\frac{1/4+\gamma^{2}}{\bar{\tau}_{\pm}^{2}}\bigg)f^{\pm}(\bar{\tau}_{\pm})=0.
\end{equation}
It is the Whittaker equation where $\gamma$ is given in Eq.~(\ref{gamma}), and $\kappa_{n}$ is defined as
\begin{equation}\label{kappa}
\kappa_{n}=\frac{\lambda k_{z}}{k_{n}}.
\end{equation}
From the definitions of $k_{n}$ and $\kappa_{n}$, given by Eqs.~(\ref{kn}) and (\ref{kappa}) respectively, the solutions to Whittaker differential equation~(\ref{Whittaker}) can be obtained following Ref.~\cite{Kobayashi:2014zza} and replacing the transverse momentum with the Landau levels, i.e., $\varepsilon_{n}=(2n+1)eB$. Therefore, the normalized positive and negative frequency solutions \cite{Bavarsad:2017oyv} that can represent the early time vacuum are, respectively,
\begin{eqnarray}
U_{\textrm{in}}(x)&=&\Omega^{-1}(\tau)\big(2k_{n}\big)^{-\frac{1}{2}}e^{-\frac{\pi}{2}\kappa_{n}}
H_{n,k_{x}}^{+}(\x_{\perp})e^{+izk_{z}}W_{i\kappa_{n},i\gamma}(\bar{\tau}_{+}), \label{uin} \\
V_{\textrm{in}}(x)&=&\Omega^{-1}(\tau)\big(2k_{n}\big)^{-\frac{1}{2}}e^{\frac{\pi}{2}\kappa_{n}}
H_{n,k_{x}}^{-}(\x_{\perp})e^{-izk_{z}}W_{i\kappa_{n},-i\gamma}(\bar{\tau}_{-}), \label{vin}
\end{eqnarray}
while those that can represent the late time vacuum are, respectively,
\begin{eqnarray}
U_{\textrm{out}}(x)&=&\Omega^{-1}(\tau)e^{+\frac{i\pi}{4}}\big(4\gamma k_{n}\big)^{-\frac{1}{2}}e^{-\frac{\pi}{2}\gamma}
H_{n,k_{x}}^{+}(\x_{\perp})e^{+izk_{z}}M_{i\kappa_{n},i\gamma}(\bar{\tau}_{+}), \label{uout} \\
V_{\textrm{out}}(x)&=&\Omega^{-1}(\tau)e^{-\frac{i\pi}{4}}\big(4\gamma k_{n}\big)^{-\frac{1}{2}}e^{-\frac{\pi}{2}\gamma}
H_{n,k_{x}}^{-}(\x_{\perp})e^{-izk_{z}}M_{i\kappa_{n},-i\gamma}(\bar{\tau}_{-}), \label{vout}
\end{eqnarray}
where the functions $W$ and $M$ are the Whittaker functions. We can use the linear transformation properties of Whittaker functions (see, for example, Ref.~\cite{Book:Olver}) to write the mode function $U_{\textrm{in}}$ in terms of the mode functions $U_{\textrm{out}}$ and $V_{\textrm{out}}$ as
\begin{equation}\label{transform}
U_{\textrm{in}}\big(x;n,k_{x},k_{z}\big)=\alpha_{n,k_{z}}U_{\textrm{out}}\big(x;n,k_{x},k_{z}\big)
+\beta_{n,k_{z}}V_{\textrm{out}}\big(x;n,-k_{x},-k_{z}\big),
\end{equation}
where the Bogoliubov coefficients are given by
\begin{eqnarray}
\alpha_{n,k_{z}}&=&e^{\frac{i\pi}{4}}\big(2\gamma\big)^{-\frac{1}{2}}\exp\Big(\frac{\pi}{2}(\gamma-\kappa_{n})\Big)
\frac{\Gamma\big(1-2i\gamma\big)}{\Gamma\big(\frac{1}{2}-i\gamma-i\kappa_{n}\big)}, \label{alpha} \\
\beta_{n,k_{z}}&=&-e^{\frac{i\pi}{4}}\big(2\gamma\big)^{-\frac{1}{2}}\exp\Big(-\frac{\pi}{2}(\gamma+\kappa_{n})\Big)
\frac{\Gamma\big(1+2i\gamma\big)}{\Gamma\big(\frac{1}{2}+i\gamma-i\kappa_{n}\big)}. \label{beta}
\end{eqnarray}
Because the momentum component perpendicular to the magnetic field direction is quantized in terms of the Landau levels and hence the quantum number $n$ as well as $k_{z}$ enters into the Klein-Gordon field equation, $\alpha_{n,k_{z}}$ and $\beta_{n,k_{z}}$ depend on $n$ and $k_{z}$. The Bogoliubov coefficients satisfy the normalization condition for bosons
\begin{equation}\label{norm}
|\alpha_{n,k_{z}}|^{2}-|\beta_{n,k_{z}}|^{2}=1.
\end{equation}
The Bogoliubov coefficient (\ref{alpha}) enables us to construct the one-loop scalar QED effective action from the Schwinger action principle.
\subsection{\label{sec:inout}Definition of the effective Lagrangian in the in-out approach}
In the in-out formalism based on the Schwinger action principle (\ref{principle}), the scattering amplitude gives an unrenormalized one-loop scalar QED effective action
\begin{equation}\label{action}
S_{\eff}=i\int dxdz\int\frac{dk_{x}}{(2\pi)}\frac{dk_{z}}{(2\pi)}\sum_{n=0}^{\infty}\log\big(\alpha_{n,k_{z}}^{\ast}\big),
\end{equation}
where the Bogoliubov coefficient $\alpha_{n,k_{z}}$ is given by Eq.~(\ref{alpha}). It follows from Eq.~(\ref{alpha}) that the spectrum is infinitely degenerate with respect to the comoving momentum component $k_{x}$. By looking back at the wave function given in Eq.~(\ref{hermit}), we see that the position of the center of the Gaussian wave function on the $y$ axis is at $k_{x}/(eB)$. Hence, the number of states \cite{Kuznetsov:2004tb} with definite comoving momentum component $k_{x}$ is given by
\begin{equation}\label{degeneracy}
\int dx\int\frac{dk_{x}}{(2\pi)}=\Big(\frac{eB}{2\pi}\Big)\int dxdy.
\end{equation}
The integral over the comoving momentum component $k_{z}$ is very challenging. As a workaround, we use a semiclassical estimate to evaluate the typical conformal time at which the pairs are created, as also done in Refs.~\cite{Frob:2014zka,Kobayashi:2014zza,Stahl:2015cra}. Under the semiclassical condition
\begin{equation}\label{semicond}
\frac{(eE)^{2}}{H^{4}}+\frac{m^{2}}{H^{2}}+12\xi\gg1,
\end{equation}
the conformal time at which a pair with comoving momentum component $k_{z}$ can be excited is roughly estimated \cite{Bavarsad:2017oyv,Bavarsad:2018lvn} to be
\begin{equation}\label{time}
\tau\sim\frac{\gamma}{k_{z}},
\end{equation}
where $k_{z}$ lies in the range $(-\infty,0)$, and corresponds to the downward tunneling transition which reduces the energy density of the vacuum \cite{Frob:2014zka}. This relation will provide a way to convert the integral over the comoving momentum component $k_{z}$ to an integral over the conformal time $\tau$. With the substitution of Eq.~(\ref{degeneracy}) into Eq.~(\ref{action}) and a change of variable using the relation (\ref{time}), we can convert this expression to the form
\begin{equation}\label{form}
S_{\eff}=i\Big(\frac{eB}{2\pi}\Big)\Big(\frac{H^{2}\gamma}{2\pi}\Big)\int\Omega^{2}(\tau)d\tau dxdydz\sum_{n=0}^{\infty}\log\big(\alpha_{n}^{\ast}\big),
\end{equation}
where the prefactor in front of the four-volume integral is the density of states. The effective Lagrangian $\Leff$ can be read directly from the definition
\begin{equation}\label{Leff:def}
S_{\eff}=\int d^{4}x\sqrt{-g}\Leff,
\end{equation}
and Eq.~(\ref{form}) gives
\begin{equation}\label{Leff:alpha}
\Leff=i\Big(\frac{e\B}{2\pi}\Big)\Big(\frac{H^{2}\gamma}{2\pi}\Big)\sum_{n=0}^{\infty}\log\big(\alpha_{n}^{\ast}\big).
\end{equation}
Substituting the Bogoliubov coefficient (\ref{alpha}), but with $-\kappa_{n}$ replaced by $\bar{\kappa}_{n}$ defined in Eq.~(\ref{gamma}), leads to
\begin{equation}\label{Leff:ln}
\Leff=i\Big(\frac{e\B}{2\pi}\Big)\Big(\frac{H^{2}\gamma}{2\pi}\Big)\sum_{n=0}^{\infty}\Big\{
\log\big[\Gamma\big(1+2i\gamma\big)\big]-\log\Big[\Gamma\Big(\frac{1}{2}+i\gamma-i\bar{\kappa}_{n}\Big)\Big]
+\frac{\pi}{2}\big(\gamma+\bar{\kappa}_{n}\big)-\frac{1}{2}\log\big(2i\gamma\big)\Big\}.
\end{equation}
By making use of the integral representation for the natural logarithm of the gamma function given by Eq.~(\ref{logamma}) and performing a Wick rotation
\begin{eqnarray}\label{lngamma}
\log\big[\Gamma\big(1+2i\gamma\big)\big]-\log\Big[\Gamma\Big(\frac{1}{2}+i\gamma-i\bar{\kappa}_{n}\Big)\Big]&=&
-i\int_{0}^{\infty}ds\bigg[\frac{e^{-2(\gamma-\bar{\kappa}_{n})s}}{2s\sin(s)}
-\frac{\cos(s)e^{-4\gamma s}}{2s\sin(s)}-(\gamma+\bar{\kappa}_{n})\frac{\cos(s)}{s}\bigg] \nn\\
&-&\frac{\pi}{2}\big(\gamma+\bar{\kappa}_{n}\big)+\frac{1}{2}\log\big(2i\gamma\big),
\end{eqnarray}
we obtain the unrenormalized effective Lagrangian
\begin{equation}\label{Leff}
\boxed{\Leff=\Big(\frac{e\B}{4\pi}\Big)\Big(\frac{H^{2}\gamma}{2\pi}\Big)\sum_{n=0}^{\infty}\int_{0}^{\infty}ds
\bigg[\frac{e^{-2(\gamma-\bar{\kappa}_{n})s}}{s\sin(s)}-\frac{\cos(s)e^{-4\gamma s}}{s\sin(s)}-2(\gamma+\bar{\kappa}_{n})\frac{\cos(s)}{s}\bigg].}
\end{equation}
The imaginary part of this expression describes the Schwinger effect in the $\dsf$ spacetime, and its real part leads to a nonlinear modification of Maxwell's theory of electrodynamics.
\subsection{\label{sec:im}Imaginary part of the effective Lagrangian and the vacuum persistence amplitude}
It can be shown that the integral in Eq.~(\ref{Leff}) converges both for $s\rightarrow0$ and $s\rightarrow\infty$. The path of integration along the positive real axis in the complex plane of the variable $s$ is deformed to infinitesimal semicircles to avoid the poles of the sine function at $s_{l}=l\pi$ with $l=1,2,3,\ldots$. The pole contributions to these infinitesimal semicircles are pure imaginary. Hence the total contribution to the imaginary part of the effective Lagrangian $i\Im\Leff$ is given by $(-i\pi)$ times the sum of the residues of the poles. That is
\begin{equation}\label{Leff:res}
\Im\Leff=\Big(\frac{e\B}{4\pi}\Big)\Big(\frac{H^{2}\gamma}{2\pi}\Big)\sum_{n=0}^{\infty}
\sum_{l=1}^{\infty}\frac{1}{l}\bigg((-1)^{l+1}e^{-2l\pi(\gamma-\bar{\kappa}_{n})}+e^{-4l\pi\gamma}\bigg).
\end{equation}
The summation over $l$ may be evaluated:
\begin{equation}\label{Leff:im}
\Im\Leff=\Big(\frac{e\B}{4\pi}\Big)\Big(\frac{H^{2}\gamma}{2\pi}\Big)\sum_{n=0}^{\infty}
\log\bigg[1+\frac{e^{-2\pi\gamma}+e^{2\pi\bar{\kappa}_{n}}}{e^{2\pi\gamma}-e^{-2\pi\gamma}}\bigg].
\end{equation}
When the pair production is small, twice the imaginary part of the effective Lagrangian $2\Im\Leff$ gives the total number of created pairs per invariant volume of the spacetime \cite{Kim:2008yt}; see Eq.~(\ref{Im:Seff}). We refer to $2\Im\Leff$ as the vacuum persistence amplitude. An expression for the imaginary part of the effective Lagrangian can be obtained directly from Eq.~(\ref{Leff:alpha}) as
\begin{equation}\label{Im:alpha}
\Im\Leff=\frac{i}{2}\big(\Leff^{\ast}-\Leff\big)=\Big(\frac{e\B}{4\pi}\Big)\Big(\frac{H^{2}\gamma}{2\pi}\Big)
\sum_{n=0}^{\infty}\log\big(1+|\beta_{n}|^{2}\big),
\end{equation}
where in the second equality we have used Eq.~(\ref{norm}). Then comparison of the right-hand sides of Eqs.~(\ref{Leff:im}) and (\ref{Im:alpha}) leads to the identification
\begin{equation}\label{logbeta}
|\beta_{n}|^{2}=\frac{e^{-2\pi\gamma}+e^{2\pi\bar{\kappa}_{n}}}{e^{2\pi\gamma}-e^{-2\pi\gamma}}.
\end{equation}
This result is reproduced by evaluating the square of the Bogoliubov coefficient (\ref{beta}), but with $-\kappa_{n}$ replaced by $\bar{\kappa}_{n}$. The quantity $|\beta_{n}|^{2}$ gives the number density of the created scalar pairs at the conformal time $\tau$ in a certain Landau level $n$. In the semiclassical regime (\ref{semicond}) the adiabatic condition is satisfied \cite{Frob:2014zka,Kobayashi:2014zza}; this implies that the particle number behaves as expected in the limit of slow expansion of the universe. Hence, under the adiabatic condition (\ref{semicond}) which implies $|\beta_{n}|^{2}\ll1$, the logarithm in Eq.~(\ref{Leff:im}) can be approximated by
\begin{equation}\label{approx}
\log\bigg[1+\frac{e^{-2\pi\gamma}+e^{2\pi\bar{\kappa}_{n}}}{e^{2\pi\gamma}-e^{-2\pi\gamma}}\bigg]\simeq
\frac{e^{-2\pi\gamma}+e^{2\pi\bar{\kappa}_{n}}}{e^{2\pi\gamma}-e^{-2\pi\gamma}}.
\end{equation}
Consequently the expression (\ref{Leff:im}) reduces to
\begin{equation}\label{im}
\Im\Leff\simeq\Big(\frac{e\B}{4\pi}\Big)\Big(\frac{H^{2}\gamma}{2\pi}\Big)\Big(\frac{1}{e^{4\pi\gamma}-1}\Big)
\sum_{n=0}^{\infty}\Big[1+e^{2\pi(\gamma+\bar{\kappa}_{n})}\Big].
\end{equation}
Summing the first term in Eq.~(\ref{im}) is obviously divergent. However, we adopt the zeta-function regularization. It will be convenient to rewrite this term as
\begin{equation}\label{zeta}
\sum_{n=0}^{\infty}1=1+\sum_{n=1}^{\infty}\frac{1}{n^{\epsilon}}=1+\zeta(\epsilon),
\end{equation}
where $\epsilon$ is a positive infinitesimal, and $\zeta(z)$ is the Riemann zeta function. Making use of the special value of the Riemann zeta function (\ref{zetazero}) to evaluate the right-hand side of Eq.~(\ref{zeta}), yields the finite result
\begin{equation}\label{zetareg}
\sum_{n=0}^{\infty}1=\frac{1}{2}.
\end{equation}
Finally applying the prescription of Eq.~(\ref{zetareg}), we arrive at a regularized expression for the imaginary part of the effective Lagrangian
\begin{equation}\label{im:final}
\boxed{\Im\Leff\simeq\Big(\frac{e\B}{4\pi}\Big)\Big(\frac{H^{2}\gamma}{2\pi}\Big)\Big(\frac{1}{e^{4\pi\gamma}-1}\Big)
\bigg[\frac{1}{2}+\sum_{n=0}^{\infty}e^{2\pi(\gamma+\bar{\kappa}_{n})}\bigg].}
\end{equation}
When the electromagnetic field is given by the vector potential (\ref{vector}), the total number of created scalar pairs per unit invariant volume of $\dsf$ spacetime reads \cite{Bavarsad:2017oyv}
\begin{equation}\label{number}
\Gamma=\frac{1}{V_{4}}\int dxdz\int\frac{dk_{x}}{(2\pi)}\frac{dk_{z}}{(2\pi)}\sum_{n=0}^{\infty}|\beta_{n,k_{z}}|^{2},
\end{equation}
where $V_{4}=\int\sqrt{-g}d^{4}x$ is the invariant volume of $\dsf$ spacetime region over which the pairs are created. The zeta-function regularization of the sum of Landau levels in Ref.~\cite{Bavarsad:2017oyv} gave Eq.~(\ref{rate}) which agrees precisely with $2\Im\Leff$ given by Eq.~(\ref{im:final}). The second term in the square bracket in Eq.~(\ref{im:final}) is the pair creation rate from the electromagnetic field while the first term is $\dsf$ spacetime radiation with the new temperature $T=m/(2\pi\gamma)$ of both the Gibbons-Hawking temperature and the Unruh temperature for the charge acceleration by the electric field \cite{Cai:2014qba} and weighted by the density of states for the electromagnetic field. We now check our result for the vacuum persistence amplitude (\ref{im:final}) by studying various relevant limiting cases.
\subsubsection{\label{sec:Hzero}Minkowski spacetime limit}
As a consistency check, it is important to verify that Eq.~(\ref{im:final}) reduces to the Schwinger effect in the Minkowski spacetime as $H\rightarrow0$. In the limit $H\rightarrow0$, $\Omega(\tau)=1$, and we use the expansion
\begin{equation}\label{expan}
\gamma-\bar{\kappa}_{n}=\frac{m^{2}}{2eE}+\frac{(2n+1)B}{2E}+\mathcal{O}(H^{2}).
\end{equation}
Inserting this expansion into Eq.~(\ref{im:final}) and then summing over the Landau levels by means of Eq.~(\ref{zetareg}), we can show that the vacuum
persistence amplitude reduces to
\begin{equation}\label{im:Hzero}
\lim_{H\rightarrow0}2\Im\Leff=\frac{e^{2}EB}{8\pi^{2}}\csch\Big(\frac{\pi B}{E}\Big)\exp\Big(-\frac{\pi m^{2}}{eE}\Big),
\end{equation}
in agreement with the well-known scalar QED Schwinger formula in Minkowski spacetime; see, e.g., Refs.~\cite{Cho:2003jb,Dunne:2004nc}.
\subsubsection{\label{sec:Ezero}Zero electric field limit}
In the limit as $E\rightarrow0$, we see from the definitions in Eq.~(\ref{gamma}) that $\bar{\kappa}_{n}=0$ and $\gamma$ reduces to
\begin{equation}\label{gamma0}
\gamma_{0}=\sqrt{\mu^{2}+12\bar{\xi}-\frac{1}{4}}.
\end{equation}
If we regularize the summation over the Landau levels by using Eq.~(\ref{zetareg}), then the vacuum persistence amplitude (\ref{im:final}) is given by the expression
\begin{equation}\label{im:Ezero}
\lim_{E\rightarrow0}2\Im\Leff=\Big(\frac{e\B H^{2}\gamma_{0}}{8\pi^{2}}\Big)\frac{1}{e^{2\pi\gamma_{0}}-1}.
\end{equation}
Therefore even if the electric field vanishes, the magnetic field affects the pair creation rate in $\dsf$ spacetime. This striking result is to be contrasted with the flat spacetime result where a purely constant magnetic field background does not produce pair \cite{Dunne:2004nc}. The fermionic case was discussed in Ref.~\cite{Tao:2024amo} in the context of the chiral anomaly.
\subsubsection{\label{sec:Bzero}Zero magnetic field limit}
In the limit of zero magnetic field we see from the definition of $\bar{\kappa}_{n}$ in Eq.~(\ref{gamma}) that it can be approximated by
\begin{equation}\label{kapprox}
\bar{\kappa}_{n}\simeq\lambda-\frac{\lambda}{2H^{2}\gamma^{2}}(2n+1)e\B,
\end{equation}
which will lead to
\begin{equation}\label{sumapprox}
\sum_{n=0}^{\infty}e^{2\pi(\gamma+\bar{\kappa}_{n})}\simeq\frac{H^{2}\gamma^{2}}{(2\pi\lambda)e\B}e^{2\pi(\gamma+\lambda)},
\end{equation}
where the summation over the Landau levels has been performed by means of Eq.~(\ref{zetareg}). Then a substitution into Eq.~(\ref{im:final}) gives
\begin{equation}\label{Bzeroapp}
2\Im\Leff=\Big(\frac{H^{2}\gamma}{2\pi}\Big)\Big(\frac{1}{e^{4\pi\gamma}-1}\Big)
\bigg[4\pi\Big(\frac{e\B}{16\pi^{2}}\Big)+\frac{H^{2}\gamma^{2}}{4\pi^{2}\lambda}e^{2\pi(\gamma+\lambda)}\bigg].
\end{equation}
The factor $e\B/(16\pi^{2})$ in the square brackets is the density of states in the perpendicular plane to the magnetic field, which should be replaced by the density of states for the electric field. To determine the density of states for the electric field, we should take account the factor of $H\gamma/2\pi$ for each of degrees of freedom in the perpendicular plane which gives $(H\gamma/2\pi)^{2}$. Therefore, in the limit as $B\rightarrow0$ the dominant contribution to the vacuum persistence amplitude (\ref{im:final}) reduces to
\begin{equation}\label{im:Bzero}
\lim_{B\rightarrow0}2\Im\Leff=\frac{H^{4}\gamma^{3}}{4\pi^{2}}\csch(2\pi\gamma)\Big(e^{-2\pi\gamma}+\frac{e^{2\pi\lambda}}{4\pi\lambda}\Big).
\end{equation}
This is in agreement with the dominant term of the pair production rate found earlier in Ref.~\cite{Kobayashi:2014zza}, in the strong electric field
regime where $\lambda\gg\mathrm{max}(1,\mu,\xi)$.
\subsubsection{\label{sec:EBzero}Zero electromagnetic field limit}
Since the limiting case of $E=0=B$ corresponds to the absence of the electromagnetic field background, we expect to recover the Gibbons-Hawking radiation
in the purely $\dsf$ spacetime from the vacuum persistence amplitude (\ref{im:final}). In this limit, $\bar{\kappa}_{n}=0$, and $\gamma$ reduces to $\gamma_{0}$ given by Eq.~(\ref{gamma0}). Furthermore, the density of states in the perpendicular plane to the magnetic field, i.e., $e\B/(16\pi^{2})$ is replaced by the density of states for the scalar field $(H\gamma_{0}/2\pi)^{2}$. If we substitute Eq.~(\ref{zetareg}) for summation over the Landau levels, then the vacuum persistence amplitude (\ref{im:final}) reduces to
\begin{equation}\label{im:EBzero}
\lim_{E,B\rightarrow0}2\Im\Leff=\Big(\frac{H^{4}\gamma_{0}^{3}}{2\pi^{2}}\Big)\frac{1}{e^{2\pi\gamma_{0}}-1}.
\end{equation}
This limiting form agrees with the particle creation rate of the real scalar field in a flat spacial section of $\dsf$ spacetime obtained earlier in Ref.~\cite{Anderson:2017hts}, up to a numerical prefactor $(0.662743)^{3}$ that depends on the details of estimation of the particle creation time.
\section{\label{sec:polar}vacuum polarization in parallel electric and magnetic fields}
Having discussed the imaginary part of the scalar QED effective Lagrangian, in this section we consider its real part. Hence we go back to the original expression (\ref{Leff}). Since we have computed the contribution of the poles of the integrand on the positive real axis arise from the sine function which leads to an imaginary part when the path of integration is deformed to infinitesimal semicircles, the remaining contribution is the principal value of the integral which is real. With this understanding concerning the integral (\ref{Leff}), the real part of the effective Lagrangian $\Re\Leff$ can be obtained from
\begin{equation}\label{Leff:pv}
\Re\Leff=\Big(\frac{e\B}{4\pi}\Big)\Big(\frac{H^{2}\gamma}{2\pi}\Big)\sum_{n=0}^{\infty}\mathcal{P}\int_{0}^{\infty}\frac{ds}{s}
\Big[\csc(s)e^{-2(\gamma-\bar{\kappa}_{n})s}-\cot(s)e^{-4\gamma s}-2(\gamma+\bar{\kappa}_{n})\cos(s)\Big],
\end{equation}
where the symbol $\mathcal{P}$ denotes the principal value. Although the integral in Eq.~(\ref{Leff:pv}) converges both for $s\rightarrow0$ and $s\rightarrow\infty$, each term in this integral is separately divergent in the $s\rightarrow0$ or ultraviolet region. Once the appropriate counterterms $\Lct$ have been subtracted from $\Re\Leff$, the remainder is finite and will be called the renormalized effective Lagrangian:
\begin{equation}\label{Lren:def}
\Lren=\Re\Leff-\Lct.
\end{equation}
It is convenient to impose renormalization conditions on $\Lren$ that could determine the set of required counterterms. In order to compare our result for the renormalized effective Lagrangian with the one-loop scalar QED effective Lagrangian in the Minkowski spacetime, known as the Weisskopf-Schwinger effective Lagrangian \cite{Weisskopf:1936hya,Schwinger:1951nm,Dunne:2004nc}, we focus on the weak curvature regime where
\begin{equation}\label{weak}
H^{2}\ll m^{2}+|eE|+|e\B|.
\end{equation}
This condition is consistent with the semiclassical condition (\ref{semicond}). To define the theory properly, we impose four renormalization conditions: \begin{enumerate}
\item The renormalized effective Lagrangian must vanish in the absence of the gravitational and electromagnetic fields \cite{Dittrich:1985yb}, which may also be interpreted as a renormalization of the cosmological constant \cite{Book:Parker}.
\item Any term in the renormalized effective Lagrangian whose structure is the same as a term in the classical Lagrangian cannot be detected individually by physical measurements and renormalizes the parameters of the theory \cite{Dittrich:1985yb}. This condition leads to the renormalization of the electromagnetic coupling constant, field strength and Newton's gravitational constant \cite{Book:Birrell}.
\item The renormalized effective Lagrangian must be an analytically varying function of the electromagnetic and gravitational fields \cite{Hollands:2001nf}.
\item The weak-field expansion of the renormalized effective Lagrangian must be a polynomial in the curvature and electromagnetic fields, which consist of terms at least of quadratic order in the electromagnetic field.
\end{enumerate}
Then, bearing these conditions, the expression
\begin{eqnarray}\label{Lct}
\Lct&=&\Big(\divp+\linp\Big)\bigg[\Big(\frac{e\B}{4\pi}\Big)\Big(\frac{H^{2}\gamma}{2\pi}\Big)\sum_{n=0}^{\infty}
\int_{0}^{\infty}\frac{ds}{s}\csc(s)e^{-2(\gamma-\bar{\kappa}_{n})s}\bigg] \nn\\
&-&\Big(\frac{e\B}{4\pi}\Big)\Big(\frac{H^{2}\gamma}{2\pi}\Big)
\sum_{n=0}^{\infty}\int_{0}^{\infty}ds\bigg[\Big(\frac{1}{s^{2}}-\frac{1}{3}\Big)e^{-4\gamma s}+2(\gamma+\bar{\kappa}_{n})\frac{\cos(s)}{s}\bigg],
\end{eqnarray}
provides the set of required counterterms that can be used to satisfy the renormalization conditions on $\Lren$ and, to remove all divergences that appear in the evaluation of the renormalized effective Lagrangian. Here, the operator $\divp$ extracts the ultraviolet divergent part of the expression and the divergences caused by $E$ in the limit $E\rightarrow0$. And, the operator $\linp$ extracts the term which has the structure $m^{2}R$. This term gives finite contribution to the renormalization of Newton's gravitational constant. It is especially convenient to represent the renormalized effective Lagrangian as
\begin{equation}\label{Lren:rewr}
\Lren=\Lone+\Ltwo,
\end{equation}
where, from Eqs.~(\ref{Leff:pv}), (\ref{Lren:def}), and (\ref{Lct}), the renormalized expressions are given by
\begin{equation}\label{L1}
\Lone=\Big(1-\divp-\linp\Big)\bigg[\Big(\frac{e\B}{4\pi}\Big)\Big(\frac{H^{2}\gamma}{2\pi}\Big)\sum_{n=0}^{\infty}\int_{0}^{\infty}
\frac{ds}{s}\csc(s)e^{-2(\gamma-\bar{\kappa}_{n})s}\bigg],
\end{equation}
and
\begin{equation}\label{L2}
\Ltwo=-\Big(\frac{e\B}{4\pi}\Big)\Big(\frac{H^{2}\gamma}{2\pi}\Big)\sum_{n=0}^{\infty}\int_{0}^{\infty}ds
\Big(\frac{\cot(s)}{s}-\frac{1}{s^{2}}+\frac{1}{3}\Big)e^{-4\gamma s}.
\end{equation}
We will show in our subsequent discussions that these expressions, which have been constructed under the set of renormalization conditions 1-4, lead to the physically reasonable results and satisfy three general properties. First, in the Minkowski spacetime limit, the renormalized one-loop effective Lagrangian reduces to the Weisskopf-Schwinger effective Lagrangian
\begin{equation}\label{proper:1}
\Lren\big(E,B\big)=\frac{1}{16\pi^{2}}\int_{0}^{\infty}\frac{ds}{s}e^{-s}\bigg[e^{2}EB\csc\Big(\frac{eEs}{m^{2}}\Big)\csch\Big(\frac{eBs}{m^{2}}\Big)
-\frac{m^{4}}{s^{2}}-\frac{e^{2}\big(E^{2}-B^{2}\big)}{6}\bigg].
\end{equation}
Second, the leading term in the asymptotic behaviour of $\Lren$ is of quadratic order in the scalar curvature for zero electromagnetic field
\begin{equation}\label{proper:2}
\Lren(R)\sim R^{2}.
\end{equation}
Three, in the weak-curvature regime (\ref{weak}), when the amplitudes of the electric and magnetic fields are well below the critical field strength $m^{2}/e$, the renormalized effective Lagrangian is a polynomial in the curvature and electromagnetic field.
\subsection{\label{sec:weak}Weak-curvature expansion}
In this subsection we will analyze the renormalized effective Lagrangian in the weak curvature regime where the criterion (\ref{weak}) is meet. Thus, we expand Eqs.~(\ref{L1}) and (\ref{L2}) around $H=0$ and work at order $H^{2}$:
\begin{eqnarray}
\gamma&=&\frac{m^{2}a}{H^{2}}+\frac{1}{2a}-\frac{H^{2}}{8m^{2}a^{3}}\Big(1-(48\bar{\xi}-1)a^{2}\Big)+\mathcal{O}(H^{4}), \label{seri:gamma}\\
\bar{\kappa}_{n}&=&\frac{m^{2}a}{H^{2}}-\frac{(2n+1)b}{2a}+\frac{H^{2}}{8m^{2}a^{3}}\Big(4(2n+1)b+3(2n+1)^{2}b^{2}\Big)
+\mathcal{O}(H^{4}), \label{seri:kappa}
\end{eqnarray}
where we have defined the dimensionless quantities
\begin{align}\label{ab}
a&=\frac{eE}{m^{2}}, & b&=\frac{e\B}{m^{2}}.
\end{align}
Substituting these expansions into the exponent of Eq.~(\ref{L1}), and expanding the resulting expression gives
\begin{equation}\label{seri:exp}
e^{-2(\gamma-\bar{\kappa}_{n})s}\simeq e^{-\frac{s}{a}}e^{-\frac{(2n+1)bs}{a}}
\Big\{1+\frac{sH^{2}}{4m^{2}a^{3}}\Big[1-(48\bar{\xi}-1)a^{2}+4(2n+1)b+3(2n+1)^{2}b^{2}\Big]\Big\}.
\end{equation}
Inserting the expansions (\ref{seri:gamma}) and (\ref{seri:exp}) into the expression (\ref{L1}), keeping terms up to order $H^{2}$, and using Eqs.~(\ref{sum1})-(\ref{sum3}) to sum over the Landau levels, we then obtain the renormalized expression
\begin{eqnarray}\label{seri:L1}
\Lone&=&\frac{m^{4}}{16\pi^{2}}\int_{0}^{\infty}\frac{ds}{s}e^{-s}
\bigg[ab\csc(as)\csch(bs)-\frac{1}{s^{2}}-\frac{(a^{2}-b^{2})}{6}\bigg] \nn\\
&+&\frac{m^{2}H^{2}}{64\pi^{2}}\int_{0}^{\infty}ds\,e^{-s}\bigg\{\big(1-48\bar{\xi}\big)
\Big(ab\csc(as)\csch(bs)-\frac{1}{s^{2}}\Big)-\frac{1}{6}-\frac{1}{s} \nn\\
&+&\Big(\frac{\csc(as)}{as}-\frac{1}{a^{2}s^{2}}\Big)\bigg[b(2+s)\csch(bs)+4b^{2}s\coth(bs)\csch(bs)
+3b^{3}s\Big(2+\sinh^{2}(bs)\Big)\csch^{3}(bs)-\frac{6}{s^{2}}\bigg]\bigg\},
\end{eqnarray}
where we have changed the variable of integration from $s$ to $as$. Note that as $H\rightarrow0$ and hence $\Omega(\tau)\rightarrow1$ where the Minkowski spacetime is recovered, $\Lone$ reduces to the Weisskopf-Schwinger effective Lagrangian represented in Eq.~(\ref{proper:1}), which is to be expected. Note also that the term proportional to $b^{3}$ in (\ref{seri:L1}) must vanish for zero magnetic field case, hence it is subtracted by the counterterm at $B=0$ to give zero contribution in this limit. Expanding the cotangent function in Eq.~(\ref{L2}) in a Taylor series about $s=0$ provides a perturbative expansion of $\Ltwo$ in the parameter $\gamma^{-1}$. By using the Taylor expansion (\ref{cot}) and integrating over $s$, we obtain
\begin{equation}\label{seri:L2}
\Ltwo=\frac{e\B H^{2}}{32\pi^{2}}\sum_{l=2}^{\infty}\frac{|B_{2l}|(2\gamma)^{2-2l}}{l(2l-1)},
\end{equation}
where Eq.~(\ref{zetareg}) has been used to regularize the summation over the Landau levels. It follows immediately from the expansion (\ref{seri:gamma}) that $\Ltwo$ is of order $H^{6}$, and can be neglected in the leading contribution to $\Lren$. Therefore, to order $H^{2}$, the renormalized one-loop effective Lagrangian is given by the expression (\ref{seri:L1}). Thus, expanding expression (\ref{seri:L1}) in a Taylor series about $s=0$ provides a perturbative expansion of $\Lren$ in the coupling constant $e$, which will give the leading order terms, i.e., the post-Maxwellian Lagrangian
\begin{equation}\label{L1:weak}
\Lren\simeq\frac{e^{4}}{5760\pi^{2}m^{4}}\Big[7\big(E^{2}-\B^{2}\big)^{2}+4E^{2}\B^{2}\Big]
+\frac{H^{2}e^{2}}{5760\pi^{2}m^{2}}\Big[\big(29-720\bar{\xi}\big)E^{2}-\big(15-720\bar{\xi}\big)\B^{2}\Big].
\end{equation}
Let the Lorentz scalar $\mathcal{F}$ and pseudoscalar $\mathcal{G}$ be defined as
\begin{align}\label{def:fg}
\mathcal{F}&=\frac{1}{4}F_{\mu\nu}F^{\mu\nu}, & \mathcal{G}&=-\frac{1}{8}\epsilon_{\mu\nu\alpha\beta}F^{\alpha\beta}F^{\mu\nu}.
\end{align}
For the electromagnetic field given by Eq.~(\ref{EBvector}), these two Lorentz-invariants are
\begin{align}\label{fg}
\mathcal{F}=\frac{1}{2}\big(\B^{2}-E^{2}\big), && \mathcal{G}=E\B.
\end{align}
The two Lorentz-invariants that are constructed from covariant derivative and covariant d'Alembertian of the electromagnetic field for the field configuration (\ref{EBvector}) yield
\begin{align}\label{nabla}
\big(\nabla_{\alpha}F_{\mu\nu}\big)^{2}&=4H^{2}\big(3\B^{2}-E^{2}\big), &
F^{\mu\nu}\Box F_{\mu\nu}&=-4H^{2}\big(2\B^{2}-E^{2}\big).
\end{align}
Equations~(\ref{fg}) and (\ref{nabla}) enable us the rewrite expression (\ref{L1:weak}) in the equivalent but more convenient Lorentz-invariant form as
\begin{equation}\label{Linvar}
\boxed{\Lren=\frac{e^{4}}{1440\pi^{2}m^{4}}\Big(7\mathcal{F}^{2}+\mathcal{G}^{2}\Big)+\frac{e^{2}}{11520\pi^{2}m^{2}}
\Big[\big(240\bar{\xi}-5\big)R\mathcal{F}+14\big(\nabla_{\alpha}F_{\mu\nu}\big)^{2}+21F^{\mu\nu}\Box F_{\mu\nu}\Big].}
\end{equation}
The advantage of this form is that it allows one to find the result in an electromagnetic field which is related to configuration (\ref{EBvector}) by a Lorentz transformation in the $\dsf$ spacetime. The one-loop effective Lagrangian in the Einstein-Maxwell background has been derived using alternative approaches worldline formalism \cite{Bastianelli:2008cu} and from more general results on heat-kernel expansion \cite{Bastianelli:2000hi}. Our result for the electromagnetic-gravity interactions in Eq.~(\ref{Linvar}) is in agreement with those results were discussed in \cite{Bastianelli:2008cu} up to the numerical coefficients.
\subsection{\label{sec:empro}Electromagnetic properties of the vacuum}
A remarkable feature of the renormalized one-loop effective Lagrangian (\ref{Lren:rewr}) is that there is a quantum nonlinearity of the electromagnetic fields that arises from virtual boson interacting with the classical fixed electromagnetic and dS backgrounds. This nonlinear feature of dS scalar QED can be expressed in terms of electric permittivity and magnetic permeability tensors. First consider the classical Maxwell Lagrangian as
\begin{align}\label{maxwel}
S_{\mathrm{Maxwell}}&=\int d^{4}x\sqrt{-g}\Lmax, & \Lmax&=-\mathcal{F},
\end{align}
where $\mathcal{F}$ is given in Eq.~(\ref{fg}). We can build the scalar QED effective Lagrangian to one-loop order by adding the one-loop correction $\Lren$ to the classical Maxwellian contribution
\begin{equation}\label{sqed}
\Lsqed=\Lmax+\Lren.
\end{equation}
The electric permittivity $\bar{\epsilon}$ and magnetic permeability $\bar{\mu}$ can be evaluated from the definitions for the electric displacement $\mathcal{D}$ and magnetic induction $\mathcal{B}$ fields \cite{Savvidy:2019grj}, respectively, via
\begin{align}\label{DB}
\mathcal{D}&=\frac{\partial\Lsqed}{\partial E}=\bar{\epsilon} E, &
\mathcal{B}&=-\frac{\partial\Lsqed}{\partial \B}=\bar{\mu} \B.
\end{align}
For simplicity, we will carry out the analysis for the approximate expression (\ref{L1:weak}). We then obtain
\begin{eqnarray}
\bar{\epsilon}&=&1+\frac{e^{4}}{1440\pi^{2}m^{4}}\Big(7E^{2}-5\B^{2}\Big)+\frac{\big(29-720\bar{\xi}\big)e^{2}H^{2}}{2880\pi^{2}m^{2}}, \label{permi} \\
\bar{\mu}&=&1+\frac{e^{4}}{1440\pi^{2}m^{4}}\Big(5E^{2}-7\B^{2}\Big)+\frac{\big(15-720\bar{\xi}\big)e^{2}H^{2}}{2880\pi^{2}m^{2}}. \label{perme}
\end{eqnarray}
These equations imply that the vacuum is birefringent where the refractive index depends on the direction of propagation; see also \cite{Banyeres:2018aax}. This phenomenon in Minkowski spacetime has been studied in detail in Refs.~\cite{Kim:2022lvn,Kim:2022fkt}. At sufficiently late times when $\B(\tau)\ll E$ is satisfied, one finds that $\bar{\epsilon}>1$ and $\bar{\mu}>1$. In this condition, the vacuum acquires: (1) A dielectric property due to the screening phenomenon \cite{Frob:2014zka} which reduces the effective electric field, and (2) a paramagnetic property that produces a magnetization in direction of the magnetic field, which in turn enhances the effective magnetic field. Consider the opposite limit: at sufficiently early times when $\B(\tau)\gg E$ is satisfied, one finds that $\bar{\epsilon}<1$ and $\bar{\mu}<1$. In this limit, the vacuum acquires: (1) an antiscreening property \cite{Frob:2014zka} which enhances the effective electric field, and (2) a diamagnetic property that produces a magnetization opposite in direction to the magnetic field, which in turn reduces the effective magnetic field. These findings signal that the electromagnetic properties of the vacuum can evolve through the expansion of the universe.
\subsection{\label{sec:ds}Effective Lagrangian in the pure $\dsf$ background}
In the absence of the electromagnetic field, $\bar{\kappa}_{n}=0$ and $\gamma$ reduces to (\ref{gamma0}). Also, the weak curvature condition (\ref{weak}) requires that $m\gg H$. Furthermore, the density of states in the perpendicular plane to the magnetic field, i.e., $e\B/(16\pi^{2})$ is replaced by $(H\gamma_{0}/2\pi)^{2}$. Then Eqs.~(\ref{Lren:rewr})-(\ref{L2}) can be written together in the simple form
\begin{equation}\label{Lren:ds}
\lim_{E,B\rightarrow0}\Lren=-\frac{H^{4}\gamma_{0}^{3}}{4\pi^{2}}\int_{0}^{\infty}ds\Big(\frac{\cot(s)}{s}-\frac{1}{s^{2}}+\frac{1}{3}\Big)
e^{-2\gamma_{0}s},
\end{equation}
where Eq.~(\ref{zetareg}) has been used to regularize the summation over the Landau levels. Expanding the cotangent function in a Taylor series about $s=0$ using Eq.~(\ref{cot}) and integrating over $s$ provides an expansion of $\Lren$ in powers of $\gamma_{0}^{-1}$ as
\begin{equation}\label{Lren:R}
\Lren=\frac{R^{2}}{576\pi^{2}}\sum_{l=2}^{\infty}\frac{|B_{2l}|\gamma_{0}^{4-2l}}{l(2l-1)}.
\end{equation}
By writing $\gamma_{0}$ in terms of the scalar curvature, expression (\ref{Lren:R}) reveals that in the pure Poincar\'{e} patch of $\dsf$ spacetime $\Lren$ admits a perturbative expansion in powers of $R$ which starts at quadratic order. In the pure global $\dsf$ spacetime, the real part of the scalar field one-loop effective action has been calculated nonperturbatively using the coincident limit of the Feynman in-in Green's function \cite{Candelas:1975du,Das:2006wg}, and using in-out formalism of Bogoliubov transformations which, in this case, vanishes in the weak curvature expansion \cite{Kim:2010cb}.
\section{\label{sec:strong}strong field expansions}
In the previous section, we mainly discussed the renormalized one-loop scalar QED effective Lagrangian in the subcritical electric and magnetic field regimes, where the parameters $a$ and $b$ are small. We shall now consider the opposite cases: the supercritical electric and magnetic fields where $a$
and $b$ are very large, and discuss the implications of the resulting gravitational corrections.
\subsection{\label{sec:Bstrng}Strong magnetic field expansion}
An important approximation to the effective Lagrangian can be obtained in a pure strong magnetic field background in the $\dsf$ spacetime. In the case of a pure magnetic field, $a$ is zero and Eq.~(\ref{seri:L1}) reduces to
\begin{eqnarray}\label{L1:B}
\Lren&=&\frac{m^{4}}{16\pi^{2}}\int_{0}^{\infty}ds\,e^{-s}\bigg\{
\Big[b\,\csch(bs)s^{-2+\epsilon}-s^{-3+\epsilon}+\frac{b^{2}}{6}s^{-1+\epsilon}\Big]
+\frac{H^{2}}{4m^{2}}\bigg[\big(1-48\bar{\xi}\big)\Big(b\,\csch(bs)s^{-1+\epsilon}-s^{-2+\epsilon}\Big) \nn\\
&+&\frac{1}{3}\Big(b\,\csch(bs)s^{\epsilon}-s^{-1+\epsilon}\Big)+\frac{1}{6}\Big(b\,\csch(bs)s-1\Big)
-\frac{2}{3}\Big(b^{2}\frac{\partial}{\partial b}\csch(bs)s^{\epsilon}+s^{-1+\epsilon}\Big) \nn\\
&+&\Big(\frac{b^{3}}{2}\frac{\partial^{2}}{\partial b^{2}}\csch(bs)s^{-1+\epsilon}-s^{-2+\epsilon}\Big)\bigg]\bigg\},
\end{eqnarray}
where we have regularized by adding a positive infinitesimal exponent $\epsilon$. And the terms involving coefficients $b^{2}$ and $b^{3}$ have been rewritten in the alternate forms which are more convenient when we do the computation. By making use of integral representations (\ref{app:gamma}) and (\ref{Hurwitz}), we can express Eq.~(\ref{L1:B}) in terms of the gamma and Hurwitz zeta functions which in the limit of $\epsilon\rightarrow0$ may be simplified to
\begin{eqnarray}\label{L1:Bsimpl}
\Lren&=&\frac{m^{4}}{96\pi^{2}}\Big[b^{2}\big(\log(2b)-1\big)-24b^{2}\zeta^{\prime}\Big(-1,\frac{1}{2}+\frac{1}{2b}\Big)-3\log(2b)-\frac{3}{2}\Big] \nn\\
&-&\frac{m^{2}H^{2}}{64\pi^{2}}\big(48\bar{\xi}-1\big)\Big[2b\zeta^{\prime}\Big(0,\frac{1}{2}+\frac{1}{2b}\Big)+\log(2b)+1\Big].
\end{eqnarray}
In the supercritical regime where $b\gg 1$ we can approximate the derivative of the Hurwitz zeta functions by making use of expansions (\ref{app:mone}) and (\ref{app:zero}). Retaining only terms that grow with $b$, we then find
\begin{eqnarray}\label{L1:Blarg}
\Lren&\simeq&\frac{1}{96\pi^{2}}\bigg[e^{2}\B^{2}\Big(\log\Big(\frac{e\B}{m^{2}}\Big)+2\log2+12\zeta^{\prime}(-1)-1\Big)
+6\log(2)m^{2}e\B-3m^{4}\log\Big(\frac{e\B}{m^{2}}\Big)\bigg] \nn\\
&+&\frac{H^{2}}{64\pi^{2}}\big(48\bar{\xi}-1\big)\bigg[\log(2)e\B-m^{2}\log\Big(\frac{e\B}{m^{2}}\Big)\bigg].
\end{eqnarray}
We are ultimately interested in the massless scalar field case for which $\Lren$ has a preferably simple form. We now proceed to calculate (\ref{L1:Blarg}) for the massless scalar field. Obviously, the massless limit $m=0$ can be taken in the gravitational corrections in (\ref{L1:Blarg}). Note, however, that the logarithmic piece diverges in the limit $m=0$, and we employ the off-shell renormalization scheme for treating this divergency. Hence, we recalculate counterterms (\ref{Lct}) at an arbitrary finite renormalization scale $M$ for massless scalar field. We eventually find that the one-loop effective Lagrangian is identical to the expression (\ref{L1:Blarg}), but with the logarithmic pieces $\log(e\B/m^{2})$ replaced by $\log(e\B/M^{2})$. The final result for the massless scalar QED effective Lagrangian is therefore
\begin{equation}\label{L1:Bmless}
\Lren=\frac{e^{2}\B^{2}}{96\pi^{2}}\Big[\log\Big(\frac{e\B}{M^{2}}\Big)+2\log2+12\zeta^{\prime}(-1)-1\Big]
+\frac{1}{64\pi^{2}}\log(2)\big(48\bar{\xi}-1\big)H^{2}e\B.
\end{equation}
In the limit $H=0$ where the redshifted magnetic field amplitude $\B$ becomes constant $B$, we observe that this result reduces to the well-known one-loop corrections to the massless scalar QED effective Lagrangian in Minkowski spacetime; see, e.g., \cite{Dittrich:1985yb}. Adding the classical Maxwell Lagrangian to (\ref{L1:Bmless}), we obtain the massless scalar QED effective Lagrangian to one-loop order
\begin{equation}\label{Lsed:B}
\Lsqed=-\frac{1}{2}\B^{2}+\frac{e^{2}\B^{2}}{96\pi^{2}}\Big[\log\Big(\frac{e\B}{M^{2}}\Big)+2\log2+12\zeta^{\prime}(-1)-1\Big]
+\frac{1}{64\pi^{2}}\log(2)\big(48\bar{\xi}-1\big)H^{2}e\B.
\end{equation}
To derive a first physical consequence of this Lagrangian, we consider the effective potential which is generally defined by \cite{Dittrich:1985yb}
\begin{equation}\label{V:def}
\V=E_{i}\frac{\partial\Lsqed}{\partial E_{i}}-\Lsqed.
\end{equation}
We can ignore the contribution of order $e^{2}$ in this substitution and obtain the effective potential in this approximation as
\begin{equation}\label{V}
\V=\frac{1}{2}\B^{2}-\frac{1}{64\pi^{2}}\log(2)\big(48\bar{\xi}-1\big)H^{2}|e\B|.
\end{equation}
Viewed as a function of $\B$, this effective potential has its minimum at
\begin{equation}\label{min}
\B_{\mathrm{min}}=\frac{1}{64\pi^{2}}\log(2)\big(48\bar{\xi}-1\big)H^{2}|e|,
\end{equation}
for $\bar{\xi}>1/48$ or equally $\xi> 3/16$. This result reveals that the phenomenon of spontaneous symmetry breaking occurs naturally in massless scalar QED theory in $\dsf$ spacetime for values of nonminimal coupling slightly larger than its value at conformal coupling. This spontaneous symmetry breaking requires the generation of magnetic fields from quantum effects in $\dsf$ spacetime. This was also remarked in Ref.~\cite{Kawati:1989vb} where the effective Lagrangian of the massless scalar QED coupled to gravity in $\dsf$ has been constructed in the functional-integral approach for two particular values of the nonminimal coupling constant: conformal coupling ($\xi=1/6$), and minimal coupling ($\xi=0$). It was subsequently shown that the minimum of the effective potential occurs close to zero in the conformal coupling case and approximately at $(1/8\pi^{2})\log(2)eH^{2}$ in the minimal coupling case. Equation~(\ref{min}) generalizes this discussion to an arbitrary values of $\xi$ and allows us to check our result for the two limiting cases of \cite{Kawati:1989vb}. In particular, from Eq.~(\ref{min}) we predict that the magnetic field will not be generated by the quantum effects for the cases $\xi=1/6$ and $\xi=0$. Thus, in the conformal coupling case ($\xi=1/6$) our prediction is in agreement with \cite{Kawati:1989vb}; whereas in the minimal coupling case ($\xi=0$) our prediction is in contradiction with \cite{Kawati:1989vb}. A conceptually similar work has also been carried out in Ref.~\cite{Kawati:1989rn} in which the symmetry behavior of the scalar field vacuum under influence of a magnetic field background in $\dsf$ spacetime has been studied. Since during the reheating phase of the Universe the potential energy stored in the inflaton field is transformed into thermal energy of the created relativistic elementary particles \cite{Kandus:2010nw}, it is physically reasonable to estimate the strength of de~Sitter-generated magnetic field (\ref{min}) at this phase. The generated magnetic field subsequently is diluted due to the expansion of the universe until today. Assuming that from reheating the entropy of the universe is conserved and considering the particle content of the standard model, the magnetic dilution factor is related to the energy scale of reheating as \cite{Kobayashi:2019uqs}
\begin{equation}\label{dilut}
\Big(\frac{\Omega_{\reh}}{\Omega_{0}}\Big)^{2}\approx\frac{10^{-43}\GeV}{9H_{\reh}},
\end{equation}
where we denote values in the universe today by the subscript 0 and at reheating by $\mathrm{reh}$. Consequently, the strength of de~Sitter-generated magnetic field (\ref{min}) at reheating is seen in the universe today
\begin{equation}\label{today}
\boxed{\B_{0}\approx 3\times 10^{-16}\Big(\xi-\frac{3}{16}\Big)\Big(\frac{eH_{\reh}}{10^{10}\GeV}\Big)~\G.}
\end{equation}
The measurements of Refs.~\cite{Essey:2010nd,Neronov:2010gir,Blunier:2025ddu} which are based on combined data observed by the Atmospheric Cherenkov Telescopes and Fermi Gamma-Ray Space Telescope indicate that the strength of the intergalactic magnetic fields must be larger than $10^{-17}\G$. On the other hand, the results derived from the analysis of Planck satellite data put an upper limit of order $10^{-9}\G$ on the strength of the intergalactic magnetic fields \cite{Planck:2015zrl}. Therefore, the prediction (\ref{today}) agrees quit well with the data in a range $10^{10}\GeV<H_{\reh}<10^{14}\GeV$ for the reheating energy scale and $\xi\sim1$. While the amplitude seems to be promising, it remains to be checked whether the backreaction of the background fields does not spoil this primordial magnetic field generation estimates. If not, this scenario, relying on the well-established Schwinger effect, is interesting as it does not require exotic initial conditions (as, e.g., in Ref.~\cite{Khalife:2024sqj}) or physics beyond the standard model. The consequences for cosmology could be numerous and diverse, ranging from solutions to the Hubble tension \cite{Jedamzik:2020krr,Jedamzik:2025cax} (see also Refs.~\cite{Stahl:2025qru,Stahl:2025czl,Lee:2025yah,Anbajagane:2025xlt} for other primordial physics impacting cosmic tensions and structure formation), to primordial magnetogenesis \cite{Stahl:2018idd,Sobol:2019xls,Kobayashi:2019uqs} with its possible impact on voids dynamics \cite{Ghosh:2025bqp}, though late time astrophysical magnetic field generation is still actively studied, for instance, the Durrive battery \cite{Durrive:2015cja}. Working out the consequences of the one-loop effective action for cosmology is beyond the scope of this work and is deferred for future studies.
\subsection{\label{sec:Estrng}Strong electric field expansion}
The renormalized one-loop effective Lagrangian for the special case of a pure electric field can be derived from Eq.~(\ref{seri:L1}) by taking the limit $b\rightarrow0$ as vanishing magnetic field. Then we can Wick-rotate the integration path by substituting $s\rightarrow-is/2a$. The result can be put into a convenient form
\begin{eqnarray}\label{L1:E}
\Lren&=&\frac{m^{2}}{64\pi^{2}}\int_{0}^{\infty}ds~e^{\frac{is}{2a}}
\bigg\{\bigg[\Big(4m^{2}+\frac{6H^{2}}{a^{2}}\Big)\Big(\frac{s}{2ia}\Big)^{-2+\epsilon}+\frac{H^{2}}{a^{2}}\Big(\frac{s}{2ia}\Big)^{-1+\epsilon}\bigg]
\bigg(\frac{1}{2}\csch\Big(\frac{s}{2}\Big)-\frac{1}{s}+\frac{s}{24}\bigg) \nn\\
&-&H^{2}\big(48\bar{\xi}-1\big)\Big(\frac{s}{2ia}\Big)^{-1+\epsilon}\Big(\frac{1}{2}\csch\Big(\frac{s}{2}\Big)-\frac{1}{s}\Big) \bigg\},
\end{eqnarray}
where $\epsilon$ is a positive infinitesimal.
In this subsection, we would like to discuss the renormalized one-loop effective Lagrangian in the supercritical electric field regime where $a\gg 1$. It is therefore appropriate to represent the proper-time integrals in Eq.~(\ref{L1:E}) by the gamma and Hurwitz zeta functions. Using formulae (\ref{app:gamma}) and (\ref{Hurwitz}) to evaluate the integrals in Eq.~(\ref{L1:E}) and talking the limit $\epsilon\rightarrow0$, we then have
\begin{eqnarray}\label{L1:Esimpl}
\Lren&=&\frac{1}{64\pi^{2}}\bigg\{\big(4e^{2}E^{2}+6m^{2}H^{2}\big)\Big[4\zeta^{\prime}\Big(-1,\frac{1}{2}-\frac{i}{2a}\Big)
-\frac{1}{6}\log\big(2ia\big)+\frac{1}{6}-\frac{1}{2a^{2}}\log\big(2ia\big)-\frac{1}{4a^{2}}\Big]-\frac{1}{6}H^{2}m^{2} \nn\\
&-&H^{2}m^{2}\Big(48\bar{\xi}-1-\frac{1}{a^{2}}\Big)\Big[\log\big(2ia\big)+2ia\zeta^{\prime}\Big(0,\frac{1}{2}-\frac{i}{2a}\Big)+1\Big] \bigg\}.
\end{eqnarray}
Equations~(\ref{app:mone}) and (\ref{app:zero}) provide appropriate expansions for the first derivative of the Hurwitz zeta functions in (\ref{L1:Esimpl}).
If we make use of these expansions and drop terms that do not grow as $E\rightarrow\infty$, we find
\begin{eqnarray}\label{L1:Elarg}
\Lren&=&-\frac{e^{2}E^{2}}{96\pi^{2}}\bigg[\log\Big(\frac{eE}{m^{2}}\Big)-1+\log4+12\zeta^{\prime}(-1)
+\frac{3m^{4}}{e^{2}E^{2}}\log\Big(\frac{eE}{m^{2}}\Big)\bigg]-\frac{1}{16\pi^{2}}\bar{\xi}Rm^{2}\log\Big(\frac{eE}{m^{2}}\Big).
\end{eqnarray}
To derive this expression, we have ignored the imaginary contribution because $\Lren$ is real. The expression in the first square brackets agrees exactly with the strong electric field expansion of the Weisskopf-Schwinger effective Lagrangian in this approximation \cite{Dunne:2004nc}, while other term is our desired gravitational corrections to the one-loop scalar QED effective Lagrangian in this limit.
\par
It is informative to explore how the coupling of the uniform electric field background with the scalar field in $\dsf$ spacetime affects the constant parameters of the theory. We begin our analysis by constructing the total effective Lagrangian. The Einstein-Hilbert Lagrangian for gravity can be defined in terms of Einstein-Hilbert action by
\begin{align}\label{LEH}
S_{\mathrm{EH}}&=\int d^{4}x\sqrt{-g}\LEH, & \LEH&=\frac{R}{16\pi G}-\rho,
\end{align}
where $G$ is Newton's gravitational constant, and $\rho$ is the vacuum energy density. Then, at one-loop, the total renormalized effective Lagrangian of the theory is constructed as a sum of Lagrangians (\ref{L1:Elarg}), (\ref{LEH}) and the classical Maxwell Lagrangian for the pure electric field case. Explicitly,
\begin{equation}\label{Ltot}
\Ltot\simeq\frac{1}{16\pi}\bigg[\frac{1}{G}-\frac{\bar{\xi}m^{2}}{\pi}\log\Big(\frac{eE}{m^{2}}\Big)\bigg]R-\rho
-\frac{m^{4}}{32\pi^{2}}\log\Big(\frac{eE}{m^{2}}\Big)+\frac{1}{2}\bigg[1-\frac{e^{2}}{48\pi^{2}}\log\Big(\frac{eE}{m^{2}}\Big)\bigg]E^{2},
\end{equation}
where we have ignored the finite constant term of order $e^{2}$ in the last square brackets. By introducing the effective scalar QED coupling constant
\begin{equation}\label{bare}
\bar{e}^{2}=\frac{e^{2}}{1-\big(e^{2}/48\pi^{2}\big)\log\big(eE/m^{2}\big)},
\end{equation}
the effective Newton's gravitational constant
\begin{equation}\label{barG}
\boxed{\bar{G}=\frac{G}{1-\big(\bar{\xi}m^{2}G/\pi\big)\log\big(eE/m^{2}\big)},}
\end{equation}
and the effective vacuum energy density
\begin{equation}\label{barrho}
\bar{\rho}=\rho+\frac{m^{4}}{32\pi^{2}}\log\Big(\frac{eE}{m^{2}}\Big),
\end{equation}
the total renormalized effective Lagrangian (\ref{Ltot}) reduces to the standard equivalent covariant form in terms of the effective parameters
\begin{equation}\label{stand}
\Ltot=\frac{1}{16\pi\bar{G}}R-\bar{\rho}-\frac{1}{4}\frac{e^{2}}{\bar{e}^{2}}F_{\mu\nu}F^{\mu\nu}.
\end{equation}
We can derive the beta functions of the parameters from
\begin{equation}\label{beta:def}
\beta_{g}(\bar{g})\equiv\frac{d\bar{g}}{d\log\big(\sqrt{eE}/m\big)},
\end{equation}
for a generic parameter $g$. Thus, Eqs.~(\ref{bare})-(\ref{barrho}) give the one-loop order expressions for the beta functions of this theory
\begin{align}\label{betas}
\beta_{e}&=\frac{e^{3}}{48\pi^{2}}, & \beta_{G}&=\frac{2\bar{\xi}m^{2}G^{2}}{\pi}, & \beta_{\rho}&=\frac{m^{4}}{16\pi^{2}}.
\end{align}
Note in particular that the result for $\beta_{e}$ is in precise agreement with the result obtained for scalar QED in Minkowski spacetime; see for instance \cite{Dunne:2004nc}. These results for $\beta_{G}$ and $\beta_{\rho}$ are consistent with the previous results obtained, for instance in  Ref.~\cite{Sola:2013gha}. The field-dependent running coupling constant $\bar{e}$ is also obtained by summing all loops in a strong magnetic field and then an electric field through the electromagnetic duality \cite{Karbstein:2019wmj}. According to Eq.~(\ref{barG}), depending on the sign of $\bar{\xi}=\xi-1/6$, three behaviors of the effective gravitational coupling $\bar{G}$ are possible in the supercritical electric field regime. If $\bar{\xi}$ is positive then $\bar{G}$ becomes large; conversely a negative $\bar{\xi}$ would imply that $\bar{G}$ becomes small. In the conformal coupling case which is $\bar{\xi}=0$, the effective gravitational coupling does not flow with the electric field strength. Returning to Eq.~(\ref{barrho}), we find that scalar QED quantum correction to the vacuum energy density increases as scalar field mass $m$ increases in the supercritical electric field regime. This quantum correction arises from the curvature independent terms in the effective Lagrangian (\ref{Ltot}), and hence it contains merely the information available in Minkowski spacetime. As clearly exposed in Ref.~\cite{Sola:2013gha}, any result that arises merely from this contribution does indeed lead to the cosmological constant fine-tuning problem and therefore it is not physically acceptable from a phenomenological point of view.
\section{\label{sec:concl}Conclusion}
Scalar pair creation and its related physics both from a strong electromagnetic field and from a dS gravitational field have been a topical issue to probe the quantum nature of spacetime, to know the inflationary response to an electromagnetic field, to constrain the electromagnetic field in the early universe and to understand better renormalization and quantum field theory in curved spacetime. In this article we have proposed a different framework to investigate the Schwinger effect in $\dsf$ spacetime: the one-loop effective action, or equivalently, the one-loop effective Lagrangian in the in-out formalism. We have derived these Lagrangians for the case of pure $\dsf$ spacetime, and the case of $\dsf$ spacetime with an electromagnetic field present. We have showed that the vacuum persistence, i.e., the imaginary part of the effective Lagrangian is in agreement with the relevant limiting cases; see Eqs.~(\ref{im:Hzero}), (\ref{im:Ezero}), (\ref{im:Bzero}), and (\ref{im:EBzero}). Furthermore, we have adopted the weak curvature condition (\ref{weak}) and analyzed the vacuum polarization (\ref{seri:L1}) which is the real part of the effective Lagrangian. In the subcritical electromagnetic field expansion, we have found the leading order correction to the one-loop effective Lagrangian in Minkowski spacetime; see Eq.~(\ref{L1:weak}). The Lorentz-invariant form of this result has been demonstrated in Eq.~(\ref{Linvar}). We have also discussed how these nonlinearities affect the electromagnetic properties of the vacuum in $\dsf$ spacetime that are characterized by the electric permittivity (\ref{permi}) and magnetic permeability (\ref{perme}). One of the most significant observations of this study is that at sufficiently early times the vacuum acquires an antiscreening property which enhances the effective electric field and a diamagnetic property whereas in the opposite limit of sufficiently late times the vacuum acquires a dielectric property due to the screening phenomenon and  paramagnetic property, the opposite behavior. Thus, we have reached the conclusion that the electromagnetic character of the vacuum can evolve through the expansion of the universe. In the absence of the electromagnetic field, as shown in Eq.~(\ref{Lren:R}) for the renormalized one-loop effective Lagrangian, we have found that it has a series in powers of $R$ starting at quadratic order.
\par
We have computed the one-loop scalar QED effective Lagrangian for pure supercritical magnetic and electric field backgrounds; the approximate results are given by Eqs.~(\ref{L1:Blarg}) and (\ref{L1:Elarg}), respectively. After examination of the scalar QED effective potential in the massless case, we have discovered that quantum effects in $\dsf$ spacetime can generate magnetic fields by spontaneous symmetry breaking mechanism if $\xi> 3/16$. This prediction is in relative agreement with the lower bound on cosmological magnetic field \cite{Blunier:2025ddu} and motivates further investigation of the Schwinger effect in that context in order to propose a full-fledged magnetogenesis scenario. Moreover, we have explored how the coupling of the uniform electric field background with the scalar field in $\dsf$ spacetime affects the fundamental constants of the theory. This analysis has illustrated that the behavior of the effective gravitational coupling constant (\ref{barG}) is characterized by the sign of the nonminimal coupling parameter $\bar{\xi}=\xi-1/6$. If $\bar{\xi}$ is positive then $\bar{G}$ becomes large in the supercritical electric field regime; conversely a negative $\bar{\xi}$ would imply that $\bar{G}$ becomes small. In the conformal coupling case $(\bar{\xi}=0)$, the effective gravitational coupling does not flow with the electric field strength. Another finding that is consistent with the previous results is that the scalar QED quantum correction to the vacuum energy density increases quartically with the scalar field mass in the supercritical electric field regime; see Eq.~(\ref{barrho}). As clearly explained in \cite{Sola:2013gha}, any result that arises merely from this contribution does indeed lead to the cosmological constant fine-tuning problem, and therefore it is not physically acceptable from a phenomenological point of view.
\par
Our new results will have implications on a fundamental level since the one-loop effective action provides a powerful tool to explore scalar QED effects for strong electromagnetic fields in a curved spacetime, here $\dsf$ spacetime. Beyond the scope of this work is to work out the primordial magnetogenesis, how primordial magnetic fields could be generated with the quantum effects described in this framework; see Refs.~\cite{Sobol:2018djj,Sobol:2019xls} for convincing steps in that direction. Other interesting physical effects due to strong background fields, such as birefringence \cite{Komatsu:2022nvu} and photon propagation and polarization, are also expected to happen in the framework we have developed here, and whether they could impact on scalar or tensor perturbations during inflation remains also an arena to be explored in the future.
\begin{acknowledgments}
The authors thank the participants of the joint program [APCTP-2025-J01] held at APCTP, Pohang, Korea for fruitful discussions. E.~B. very much appreciates the support of the University of Kashan, Grant No.~1391830/1. S.~P.~K. was supported in part by the Institute of Basic Science (Grant No.~IBSR038-D1) and also by National Research Foundation of Korea (NRF) funded by the Ministry of Education (2019R1I1A3A01063183), and thanks Rong-Gen Cai for the warm hospitality at the Institute of Theoretical Physics (ITP), Chinese Academy of Sciences (CAS) and Ningbo University.
\end{acknowledgments}
\appendix*
\section{\label{app:A}Reference mathematical formulae}
This Appendix collects together some the mathematical formulae that are used in this article. The summations over the Landau levels can be calculated with the help of
\begin{eqnarray}
\sum_{n=0}^{\infty}e^{-(2n+1)z}&=&\frac{1}{2}\csch(z), \label{sum1} \\
\sum_{n=0}^{\infty}(2n+1)e^{-(2n+1)z}&=&\frac{1}{2}\coth(z)\csch(z), \label{sum2} \\
\sum_{n=0}^{\infty}(2n+1)^{2}e^{-(2n+1)z}&=&\frac{1}{2}\csch(z)+\csch^{3}(z). \label{sum3}
\end{eqnarray}
Here, Eqs.~(\ref{sum2}) and (\ref{sum3}) are obtained by differentiating (\ref{sum1} ) with respect to $z$. One needs the Laurent series of the functions cosecant and cotangent around the origin
\begin{eqnarray}
\csc(z)&=&\frac{1}{z}+\sum_{n=1}^{\infty}\frac{(2^{2n}-2)|B_{2n}|}{(2n)!}z^{2n-1}, \label{csc} \\
\cot(z)&=&\frac{1}{z}-\sum_{n=1}^{\infty}\frac{2^{2n}|B_{2n}|}{(2n)!}z^{2n-1}, \label{cot}
\end{eqnarray}
where $B_{n}$ are the Bernoulli numbers \cite{Gradshteyn:1943cpj}. The first few nonzero Bernoulli numbers are
\begin{align}\label{Bernoulli}
B_{0}&=1, & B_{1}&=-\frac{1}{2}, & B_{2}&=\frac{1}{6}, & B_{4}&=-\frac{1}{30}, & B_{6}&=\frac{1}{42}, & B_{8}&=-\frac{1}{30}.
\end{align}
The gamma function can be expressed as a definite integral
\begin{align}\label{app:gamma}
\Gamma(z)=\int_{0}^{\infty}e^{-s}s^{z-1}ds, && \Re(z)>0.
\end{align}
Since we have chosen to represent the effective Lagrangian by a proper-time integral, it is convenient to use the following \cite{Gradshteyn:1943cpj} integral representation for the natural logarithm of the gamma function:
\begin{align}\label{logamma}
\log\Gamma(z)=\int_{0}^{\infty}\bigg[\frac{e^{-zs}-e^{-s}}{1-e^{-s}}+(z-1)e^{-s}\bigg]\frac{ds}{s}, && \Re(z)>0,
\end{align}
and an analytical continuation will be used for $\Re(z)\leq 0$. The Riemann zeta function is classically defined by
\begin{align}\label{app:zeta}
\zeta(z)=\sum_{n=1}^{\infty}\frac{1}{n^{z}}, && \Re(z)>1.
\end{align}
The integral representation of the Riemann zeta function shows that it can be analytically continued \cite{Elizalde:1994gf}, and yields the particular value
\begin{equation}\label{zetazero}
\zeta(0)=-\frac{1}{2}.
\end{equation}
One of the integral representations of the Hurwitz zeta function is
\begin{align}\label{Hurwitz}
\zeta(z,a)=\frac{1}{\Gamma(z)}\int_{0}^{\infty}ds\,s^{z-1}\frac{e^{-as}}{1-e^{-s}}, && \Re(z)>1,~\Re(a)>0,
\end{align}
and it can be analytically continued to other values of $z$ and $a$. The first derivative of the Hurwitz zeta function is defined by
\begin{align}\label{DHurwit}
\zeta^{\prime}(-n,a)=\frac{\partial\zeta(z,a)}{\partial z}\Bigg|_{z=-n}, && n=0,1,2,\ldots.
\end{align}
The asymptotic expansion
\begin{equation}\label{app:mone}
\zeta^{\prime}\Big(-1,\frac{1}{2}+z\Big)=\frac{1}{2}\zeta(-1)\log2-\frac{1}{2}\zeta^{\prime}(-1)-\frac{z}{2}\log2
+\big(1-\log4-\gamma_{\mathrm{E}}\big)\frac{z^{2}}{2}+\sum_{n=2}^{\infty}\frac{(-1)^{n}(2^{n}-1)\zeta(n)}{n(n+1)}z^{n+1},
\end{equation}
is valid for $|z|<1$, the details of the derivation is worked out in Ref.~\cite{Dunne:2004nc}. Here, $\gamma_{\mathrm{E}}$ is the Euler-Mascheroni constant. The numerical values of the Riemann zeta function and its derivative at the point -1 are \cite{Elizalde:1994gf}
\begin{align}\label{zetamone}
\zeta(-1)=-\frac{1}{12}, && \zeta^{\prime}(-1)=-0.16542115\ldots.
\end{align}
Following the same procedure used to obtain Eq.~(\ref{app:mone}), one can show that for $|z|<1$,
\begin{equation}\label{app:zero}
\zeta^{\prime}\Big(0,\frac{1}{2}+z\Big)=-\frac{1}{2}\log2-\big(\log4+\gamma_{\mathrm{E}}\big)z
+\sum_{n=2}^{\infty}\frac{(-1)^{n}(2^{n}-1)\zeta(n)}{n}z^{n}.
\end{equation}

\end{document}